\documentclass[aps,pre]{revtex4}
\usepackage{graphicx}
\usepackage{color}

\def\be{\begin{equation}}
\def\ee{\end{equation}}     
\def\bfi{\begin{figure}}
\def\efi{\end{figure}}
\def\bea{\begin{eqnarray}}
\def\eea{\end{eqnarray}}

\begin{document}

\title{Phase ordering in 3d disordered systems}

\author{Federico Corberi}
\affiliation {Dipartimento di Fisica ``E.~R. Caianiello'', and INFN, Gruppo Collegato di Salerno, and CNISM, Unit\`a di Salerno,Universit\`a  di Salerno, 
via Giovanni Paolo II 132, 84084 Fisciano (SA), Italy.}

\author{Eugenio Lippiello}
\affiliation{Dipartimento di Matematica e Fisica, Seconda Universit\`a di Napoli,
Viale Lincoln 5, 81100 Caserta, Italy.}

\author{Marco Zannetti}
\affiliation {Dipartimento di Fisica ``E.~R. Caianiello'', and INFN, Gruppo Collegato di Salerno, and CNISM, Unit\`a di Salerno,Universit\`a  di Salerno, via Giovanni Paolo II 132, 84084 Fisciano (SA), Italy.}

\begin{abstract}

We study numerically the phase-ordering kinetics of the site-diluted and bond-diluted Ising models
after a quench from an infinite to a low temperature. 
We show that the speed of growth of the ordered domain's size is non-monotonous with
respect to the amount of dilution $D$: Starting from the pure case $D=0$ the system slows
down when dilution is added, as it is usually expected when disorder is introduced, 
but only up to a certain value $D^*$ beyond which the speed of growth raises again.
We interpret this counterintuitive fact in a renormalization-group inspired framework,
along the same lines proposed for the corresponding two-dimensional systems,
where a similar pattern was observed.

\end{abstract} 

\maketitle

\section{Introduction}

The quench of systems, like ferromagnets or binary mixtures, to
below the critical point is characterized by the formation and subsequent growth of domains. The typical feature of 
the phase ordering process is dynamical scaling, whereby physical properties become time-independent if lengths are measured
in units of $L$, i.e. the typical domain size at time 
$t$~\cite{bray,2dinoirf,zan,BCKM,ioeleti,crp,rev}. This scaling scenario is 
expected to be valid also in presence of quenched disorder when it does not prevent the phase-ordering process \cite{crp,Puri}. 
In the case of pure systems power law growth $L \sim t^{1/z}$ is well established. Conversely,  
when disorder is present the nature of asymptotic growht has been 
much debated~\cite{diluted,nuovo,puripowerlaw,hp06,
lastparma,decandia,EPL,CLMPZ,variousnoi2,various2,10noirf,pp,HH,parma}. 

In general terms, disorder generated energy barriers bring about a slowing 
down of the coarsening process~\cite{crp,pcp91,bh91,puripowerlaw,hp06,sab08,various2,decandia,EPL,CLMPZ,variousnoi2,pp} and,
as a rule of thumb, more disorder means slower growth.
However, recent studies have shown that the growth low does not
behave monotonically with increasing disorder in 
the two dimensional site or bond diluted Ising model~\cite{diluted}.
In these systems, where 
a fraction $D$ of sites or bonds is randomly removed from a regular lattice,    one observes 
that, for sufficiently large $D$,
more disorder produces faster growth. 
More precisely, starting from the 
pure case with $D = 0$, where the usual temperature-independent
power-law $L\sim t^{1/2}$ is obeyed, one enters a region with an 
asymptotic logarithmic behavior as soon as $D > 0$. However
a temperature-dependent power-law growth
is observed right at $D=D_c=1-p_c$, where $p_c$ is the critical 
percolation density, thus leading to a non-monotonic
dependence of the speed of growth upon the disorder strength $D$.
This counterintuitive behavior has been interpreted in terms of the topology of the network of occupied sites or bonds~\cite{diluted,nuovo}
with the temperature-dependent algebraic behavior
associated to the fractal properties of the percolating network.

In this Article, by studying numerically the site-diluted Ising model
(SDIM) and the bond diluted one (BDIM)
in $d=3$ we find a non-monotonous dependence of $L$ on disorder 
qualitatively similar to the two-dimensional case.
This suggests that the relation between growth-law and topology
might be a rather generic property, a fact that could provide new insights into
the problem of phase-separation in disordered ferromagnets.

This paper is organized as follows. In Sec. \ref{Models} we describe the 
models, the SDIM and the BDIM,  that will be considered and studied throughout this paper. In Sec. \ref{GL} we present an overview of the results in $d=2$ whereas numerical simulations of the $d=3$ models are discussed in Sec. \ref{3d}.
In Sec. \ref{CONCL} we conclude the paper with a final
discussion of the results and of some open issues.

\section{The diluted Ising Model}
\label{Models}

\subsection{The network}

In the SDIM a network is prepared by generating configurations of occupied sites as in random
percolation. On the sites of a square three-dimensional lattice there are independent
random variables $n_i$, which take values $n_i=1$ (occupied site) with probability $p$
and $n_i=0$ (empty site) with probability $D=1-p$. The substrate is formed by the set of
occupied sites. Any couple of
nearest-neighbour occupied sites is connected by a {\it bond}.

In the BDIM, instead, starting from a regular square three-dimensional lattice
one removes bonds between nearest-neighbouring sites $<ij>$ at random with probability $D$.
Specifically, an independent random variable $J_{ij}$ is defined on each couple of nearest sites,  
which takes values $J_{ij}=0$ with probability $D$ (bond is absent) and $J_{ij}=J_0$ with
probability $1-D$ (bond is present). 

For both models, $D$ will be referred to as dilution.
We now describe the geometrical structure of the network as $D$
goes from high to low values. Recalling that 
the percolation threshold is 
$p_c\simeq 0.3116$ and $p_c\simeq 0.2488$ for site and bond percolation respectively, 
for $D> D_c = 1-p_c$, the network is a patchwork of finite clusters of occupied sites (bonds)
with a characteristic size that diverges at the threshold
and a fractal dimension $d_f\simeq 2.5$ \cite{Stauffer}. 
The dilution range $D>D_c$ will not be 
considered in this paper since, lacking an infinite cluster, 
perpetual coarsening cannot be 
sustained. At $D_c$, the size of the percolating cluster 
of sites (bonds) diverges. 
As $D$ is lowered below $D_c$, the infinite cluster has a fractal geometry 
over distances up to a typical length $\xi(D)$ that diverges at the threshold as
\be
\xi (D)\sim (D_c-D)^{-\nu},
\label{divxi}
\ee
with $\nu\simeq 0.875$,
whereas it becomes compact over larger distances. Moreover, clusters of finite size 
are also present. 
The structure of the network described above holds in the range $D_c \geq D \geq D^*$, 
where $D^*$ is a dilution value where
\be
\xi(D^*) = a ,
\label{sub.3}
\ee
and $a$ is a microscopic length, like the lattice spacing. 
For $D<D^*$, 
the infinite cluster is compact on any lengthscale and finite clusters are absent. 
The remaining far-apart vacancies inside this cluster are 
essentially single-site objects. 
Their average distance defines a new characteristic length 
\be
\lambda(D) = a(D^*/D)^{\frac{1}{d}}. 
\label{sub.4}
\ee

\subsection{Spin system}

The diluted Ising model is
described by the following Hamiltonian
\be
H=-\sum _{<ij>}K_{ij}\sigma _i \sigma _j.
\label{ham}
\ee
Here $\sigma _i=\pm 1$ are spins on the 3-d square lattice, and the couplings 
$K_{ij}$ for the SDIM is chosen as $K_{ij}=J_0n_in_j$, where $J_0$ is a constant,
and $n_i=0$ or $n_i=1$ with probability $D$ and $1-D$, respectively.
For the BDIM, instead, $n_i=1$ on all sites and $K_{ij}=0$ or $K_{ij}=J_{o}$  with probability $D$ and $1-D$, respectively.
Notice that, since $J_{0}> 0$, the model is ferromagnetic.

\subsection{Equilibrium states}

\begin{figure}[t]
	\vspace{1cm}
    \centering
   \rotatebox{0}{\resizebox{.5\textwidth}{!}{\includegraphics{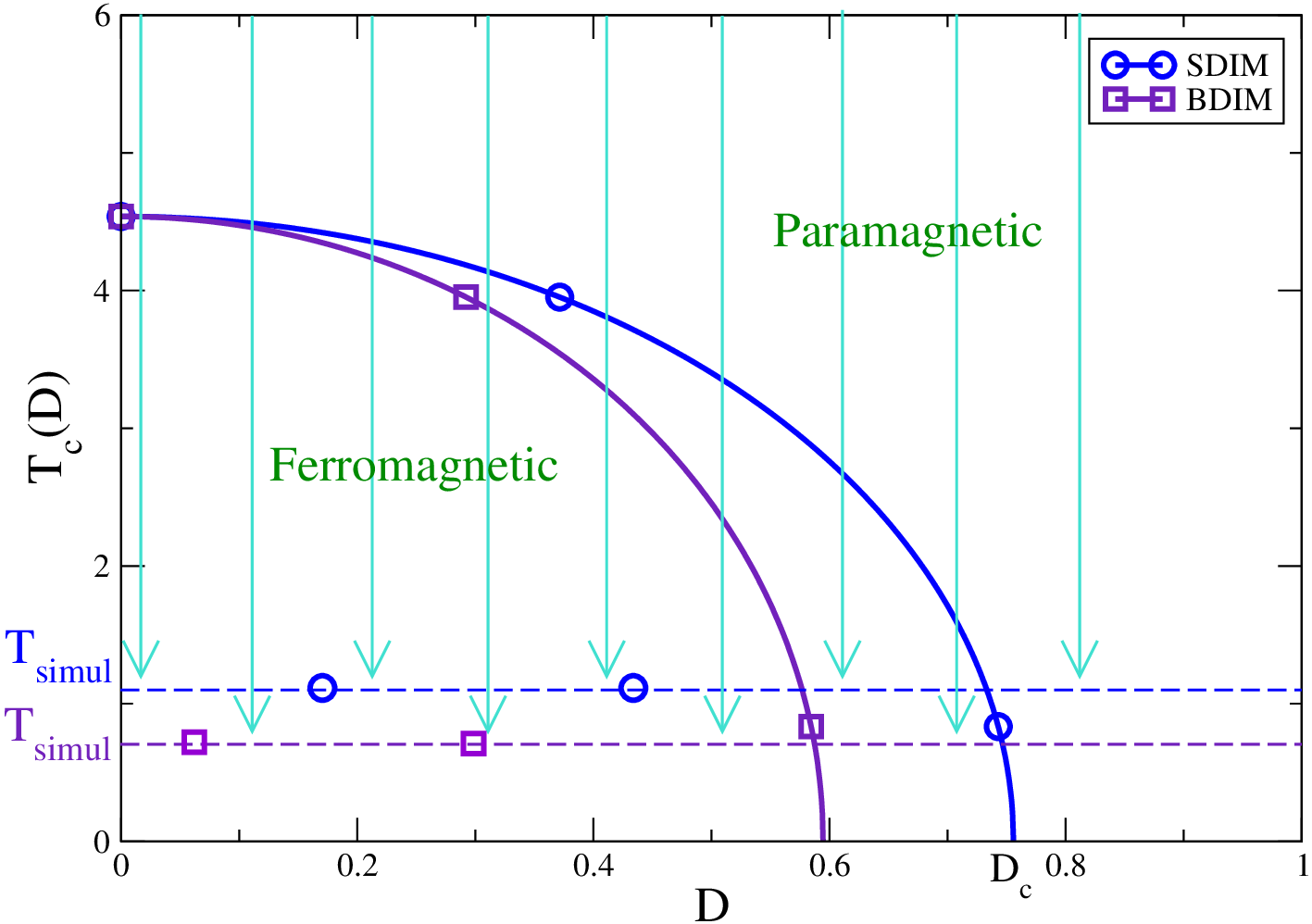}}}
   \caption{(Color online) Pictorial representation of the equilibrium phase diagram of the 
diluted Ising model in $d=3$. The bold blue line with circles and the bold indigo one with squares are the critical temperature $T_c(D)$ of the SDIM and of the BDIM, respectively.
   The temperatures $T=T_{simul}$ where simulations will be performed
   are marked with horizontal dashed lines, blue with circles for the SDIM and indigo with squares for the BDIM. Vertical arrows represent the quenching processes
   considered in the simulations.}
\label{fig_Tc}
\end{figure}

The equilibrium phase-diagram of the model is pictorially represented in 
Fig. \ref{fig_Tc}. 
For $D \leq D_c$, at low enough temperature the system exhibits ferromagnetic order. 
In the $(D,T)$ plane there is a critical line $T_c(D)$, which separates the paramagnetic from 
the ferromagnetic phase. 
The critical temperature $T_c(D)$ , which in the following will be measured in units 
of $k_B/J$, where $k_B$ is the Boltzmann constant, decreases from the pure value 
($T_c(0) \simeq 4.5$) as the dilution is increased and vanishes at $D_c$ ($T_c(D_c) = 0$).

\subsection{Time Evolution}
\label{timev}

Non-conserved dynamics\cite{bray,2dinoirf} is implemented by evolving the spins 
with single-spin-flip transition rates of the Glauber form
\be
w(\sigma _i \to -\sigma _i)=\frac{1}{2} \left [1-\sigma _i \tanh (H_i^W/T)\right ] .
\label{transr}
\ee
Here, $H_i^W$ is the local Weiss field obtained by the sum
\be
H_i^W = \sum _{j \in L_i} K_{ij} \sigma_j
\ee
over the set of nearest-neighbors $L_i$ of $i$. 

The system is initially prepared in the infinite-temperature disordered state and, at the time $t=0$, 
it is suddenly quenched 
to a finite temperature $T$. This temperature is set 
in the simulations to a value
$T_{simul}=1.1$ for the SDIM and $T_{simul}=0.75$ for the BDIM. These values are chosen as a compromise:
on the one hand we want to work in the limit of low-temperatures because the interplay between
the two types of growth-law we are interested in is expected in the $T\to 0$ limit \cite{diluted}.
On the other hand very low temperatures are useless in simulations for at least a couple
of reasons. The first is that at very low temperatures pinning effects
make the dynamics very slow. As a result the growing length 
$L(t,D)$ does not vary appreciably in the time range accessed
by the simulations and it is difficult to discriminate between different growth-laws.
The second related reason is that at low temperatures the pinning centres produce a stop and go
behavior of interfaces which, in turn, gives rise to an oscillatory behavior on top
of the growth of $L(t,D)$.
These oscillations prevent a reliable determination of the growth-law.
A similar behavior is observed on fractal substrates \cite{lastparma,parma,nuovo}

In Sec. \ref{3d} we will present the results of  simulations of systems with different values 
of $D$ in the range $[0,D_c]$. 
It must be noticed that, as discussed more detailedly in \cite{diluted}, for dilutions $D=D_c$
or close to it the working temperature $T=T_{simul}$ is necessarily larger than $T_c(D)$, see Fig. \ref{fig_Tc}. 
Hence we are quenching above the critical temperature $T_c(D_c)=0$, the system will eventually 
relax to a disordered state with a finite spin coherence length $\xi _\sigma(T)$ 
(not to be confused with the substrate property $\xi (D)$) and
the growth of $L(t,D)$ will eventually stop at 
$t=t_{eq}(T)$ when $L(t,D)\simeq \xi _\sigma(T)$. 
In a preasymptotic stage with $L(t,D)\ll \xi _\sigma(T)$ coarsening takes place and,
since $\xi _\sigma(T)$ diverges very fast for $T\to 0$, this can last for a very long time. 
The situation is similar to the one
found in other ferromagnetic systems with $T_c=0$ as, for instance, the one-dimensional 
Ising model  \cite{lip,Claudio}. We have checked that  
$T_{simul}$ is sufficiently low as to never observe the equilibration of the system in the 
range of simulated times, and that the system keeps on coarsening at any 
simulated time. 

\subsection{Observables}

The observable quantity of interest in this paper is the typical domain size $L(t,D)$.
For a model defined on a disconnected network, as it happens for $D>D^*$, phase 
ordering occurs independently on the various parts of the system and, 
correspondingly, different definitions of the growing length can be given. 
Following \cite{diluted,nuovo}, we determine 
this quantity from the inverse excess energy as 
\be
L(t,D)=[E(t,D)-E_\infty(D)]^{-1} ,
\label{lt}
\ee
where $E(t,D)=\langle H\rangle$ is the energy at time $t$, the angular brackets denote a 
non-equilibrium ensemble average, taken over random initial conditions and over dynamical trajectories and disorder realizations,
and $E_\infty (D)$ is the energy of the equilibrium state at the final temperature $T$. 
Eq.~(\ref{lt}), which is often used to determine $L(t,D)$ in non-diluted systems
\cite{bray}, has, in the present case, the further advantage that the disconnected finite parts of the substrate which are already 
ordered do not contribute to the computation of $L(t,D)$. 
Indeed each droplet is surrounded by empty sites in the SDIM or by null bonds in the BDIM
and hence there is no excess 
energy associated with it when all spins are aligned, as discussed in 
\cite{diluted,nuovo}. 
Using Eq.(\ref{lt}), then, one computes $L(t,D)$ only on the regions where growth is active,
which is the kind of information we are interested in.

\section{Growth Law}
\label{GL}

\subsection{Overview of the behaviour of the $2d$-SDIM} \label{over}

In \cite{diluted,nuovo} it was shown that
the kinetic properties of the $2d$-SDIM and BDIM are due to the presence of three fixed points, 
in the sense of the renormalization group, existing on the $D$ axis. 
The first fixed point is the trivial
one of the pure system, located at $D=0$, and is associated to the usual
power-law growth $L(t,0)\sim t^{1/2}$. The second one is the {\it percolative}
fixed point located at $D_c$ and is characterized by a temperature-dependent
power-law growth
\be
L(t,D_c)\sim t^{1/\zeta (T)},
\label{ldc}
\ee
with $\zeta (T)\ge 2$. 
This new power-law growth is due to the fact that diluted sites act as pinning
centres where interfaces get stucked unless an activation energy $\Delta E$ is supplied by
the thermal bath. According to the argument presented
in \cite{diluted,nuovo} this energy scales as $\Delta E\sim A \ln L(t,D)$.
This is expected to be true for sufficiently low temperatures.
Assuming an Arrhenius time $t\sim \exp [\Delta E/(k_BT)]$ to exceed such energetic 
barriers one obtains Eq. (\ref{ldc}) with an exponent
\be
1/\zeta(T)=k_BT/A
\label{expT}
\ee
proportional to the absolute temperature.

The two fixed points discussed above are repulsive,
in the sense that as soon as $D\neq 0$, or $D\neq D_c$ the asymptotic dynamics
is governed by a different fixed point, located at $D=D^*$ (roughly half way
between $D=0$ and $D=D_c$), 
to which a logarithmic increase of $L(t,D^*)$ is associated. Due to these fixed point 
structure, if the system is prepared with an intermediate dilution 
between $D=0$ and $D^*$ (or equivalently between $D^*$ and $D_c$)
a crossover pattern is observed at a certain time $t_{cross}(T,D)$ from an initial transient regime 
governed by the nearest unstable fixed point ($D=0$ or $D_c$ respectively),
with a power-law increase of the domains size, to a late regime controlled
by the attractive point at $D^*$, characterized by a logarithmic $L(t,D)$.
The crossover phenomenon occurs when $L(t,D)$ reaches the typical size 
associated to dilution, which is either $\lambda (D)$ or $\xi (D)$ depending
on $0<D<D^*$ or $D^*<D<D_c$. 
Given that both $\lambda$ and $\xi$ decrease approaching $D^*$ 
(Eqs. (\ref{divxi},\ref{sub.4})), the slowest possible growth, namely the 
one where $t_{cross}(T,D)$ is
smaller, is obtained at $D=D^*$. Therefore 
by comparing $L(t,D)$ for different values of $D$ one finds a non-monotonous 
behavior: The growth
slows down in going from $D=0$ to $D=D^*$ and then speeds up again
when $D$ is further increased from $D=D^*$ up to $D=D_c$. 
This can represent a practical tool to identify $D^*$.

In the next section we will show that a similar structure is observed also
in $3d$.

\section{Numerical results}
\label{3d}
\subsubsection{Simulation details}

The details of the simulations are as follows. We have considered a three-dimensional square lattice 
system of $N={\cal L}^2$ sites, with ${\cal L}=240$ for the SDIM and ${\cal L}=150$
for the BDIM. We have checked that, with this choice, no finite-size effects can be detected 
in the time-regime accessed by the simulations. For every choice of the parameters, 
we have performed a certain number (in the range 10-100) of independent runs 
with different initial 
conditions, disorder realizations and thermal histories in order to populate the non-equilibrium ensemble needed to 
extract average quantities. 

To speed up the simulations we have used a modified dynamics where 
spin flips in the bulk of domains, namely those aligned with all the nearest neighbors, 
are prevented. This modified dynamics does not alter the behavior of the quantities we are 
interested in, as has been tested in the $2d$ version of the  models and in a large number of 
different cases \cite{nobulk}. We have checked that this is also true in the present study.

\subsubsection{3d results: SDIM}

\begin{figure}[t]
    \centering
    \vspace{1cm}
   \rotatebox{0}{\resizebox{.6\textwidth}{!}{\includegraphics{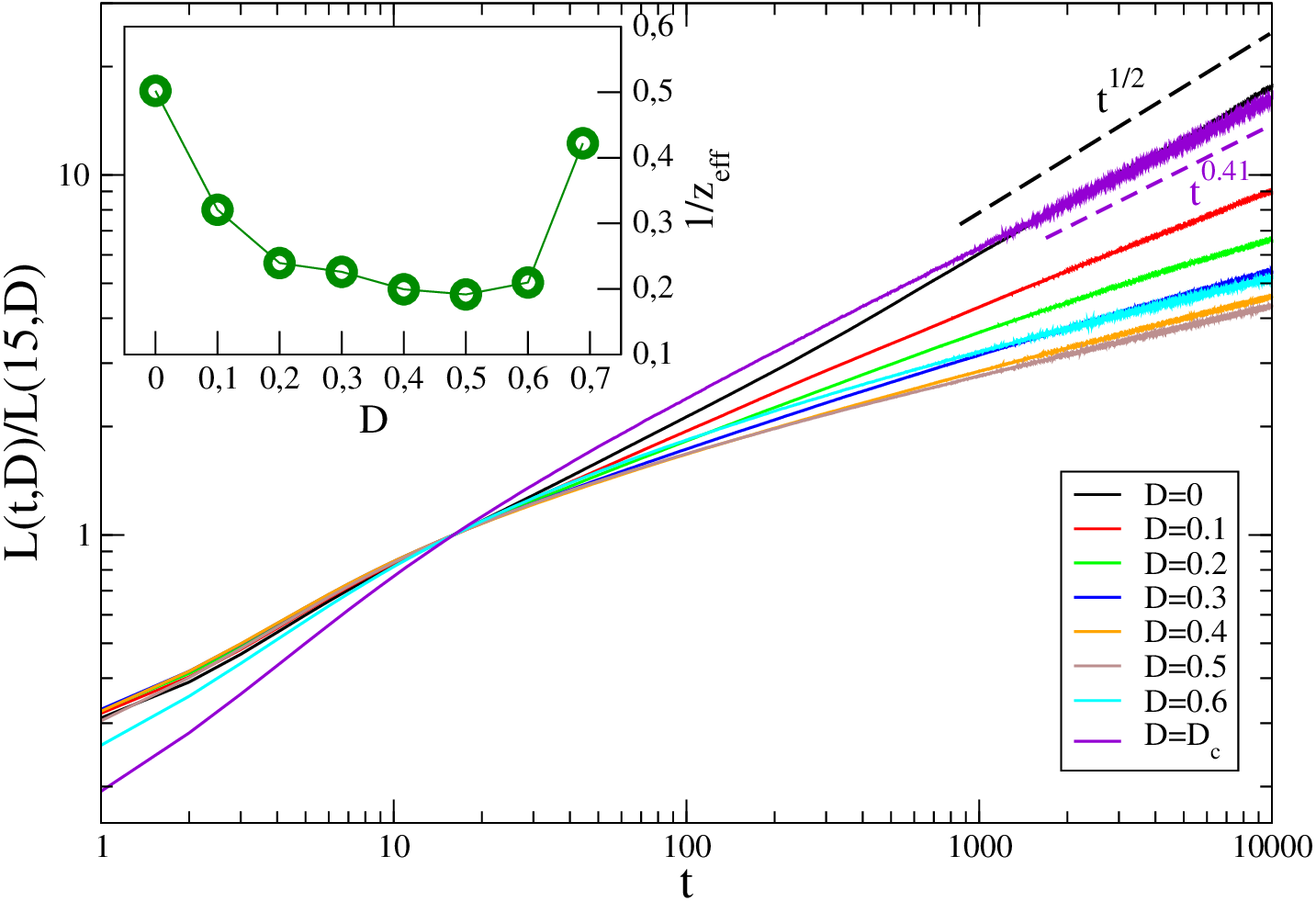}}}
      \caption{(Color online) The behavior of the typical length $L(t,D)/L(15,D)$ after a quench to 
$T=T_{simul}$ is shown for the SDIM 
   for several dilutions $D$ (see legend) in a double logarithmic plot. 
   The dashed-black line is the pure growth-law $L(t,0)\sim t^{1/2}$ and the dashed-violet one is the
   power-law $L(t,D_c)\sim t^{0.41}$. 
In the inset the effective exponent $1/z _{eff}$ (see text) is plotted
against $D$.}
\label{fig1}
\end{figure}

Let us start by considering the SDIM.
The time dependence of $L(t,D)$, for various dilution values,
is plotted in Fig.~\ref{fig1} for the $3d$-SDIM.
Notice that we plot $L(t,D)/L(15,D)$
in order to normalize the curves at the microscopic time 
$t_{micro}\simeq 15$ when the late-stage scaling regime is entered.   

In the pure case the asymptotic power-law $L(t,0)\sim t^{1/2}$ is expected. 
In the relatively small range of simulated times we measure
a somewhat smaller exponent (in the last decade a best fit procedure yields $1/z\simeq 0.46$, 
as already reported in 
\cite{altriIsing,noiIsing}, because the convergence
to the true asymptotic behavior is very slow, as explained in \cite{noiIsing}.
Indeed the observed {\it effective} exponent $1/z$ appears to increase with time. However here we are mainly interested in the non-monotonic behavior
of the speed of growth for which an accurate estimate of $z$ is not necessary.
In fact, it is clearly observed that, by increasing $D$, the growth law becomes slower and slower  
up to a certain value of $D\simeq 0.5$ that, following 
\cite{diluted}, we identify with $D^*$.
For $D>D^*$  growth becomes faster upon increasing $D$
up to $D_c$. This behavior results in a non-monotonic $D$-dependence of $L(t,D)$, at fixed time $t$. This can be clearly appreciated in the inset of 
Fig. \ref{fig1} where the behavior of the effective exponent obtained
by fitting the data for various $D$ with a power-law (regardless of the
fact that the actual growth-law might not be power-law) in the
time-range $t\ge 5000$.

Having established the non-monotonous dependence of the speed of
growth on $D$, we make some comments on the form of the growth-law.
At the dilution $D=D^*$ we observe a  downwards banding of $L(t,D)$ which indicates that
the growth is slower than algebraic, suggesting in this case 
a logarithmic law. 
We recall that a form
$L(t,D)\simeq (\ln t)^{1/\phi}$ was predicted in a model with 
bond disorder in \cite{HH}, which, however,
due its slow character has never been accurately demonstrated, 
to the best of our knowledge, 
even in the less numerically demanding 2d cases.
Clearly in this $3d$ case the task is much more demanding, and the
time range accessible in the simulations is not sufficient for
a precise determination of the analytical form of the growth-law.

\begin{figure}[t]
    \centering
    \vspace{1cm}
   \rotatebox{0}{\resizebox{.75\textwidth}{!}{\includegraphics{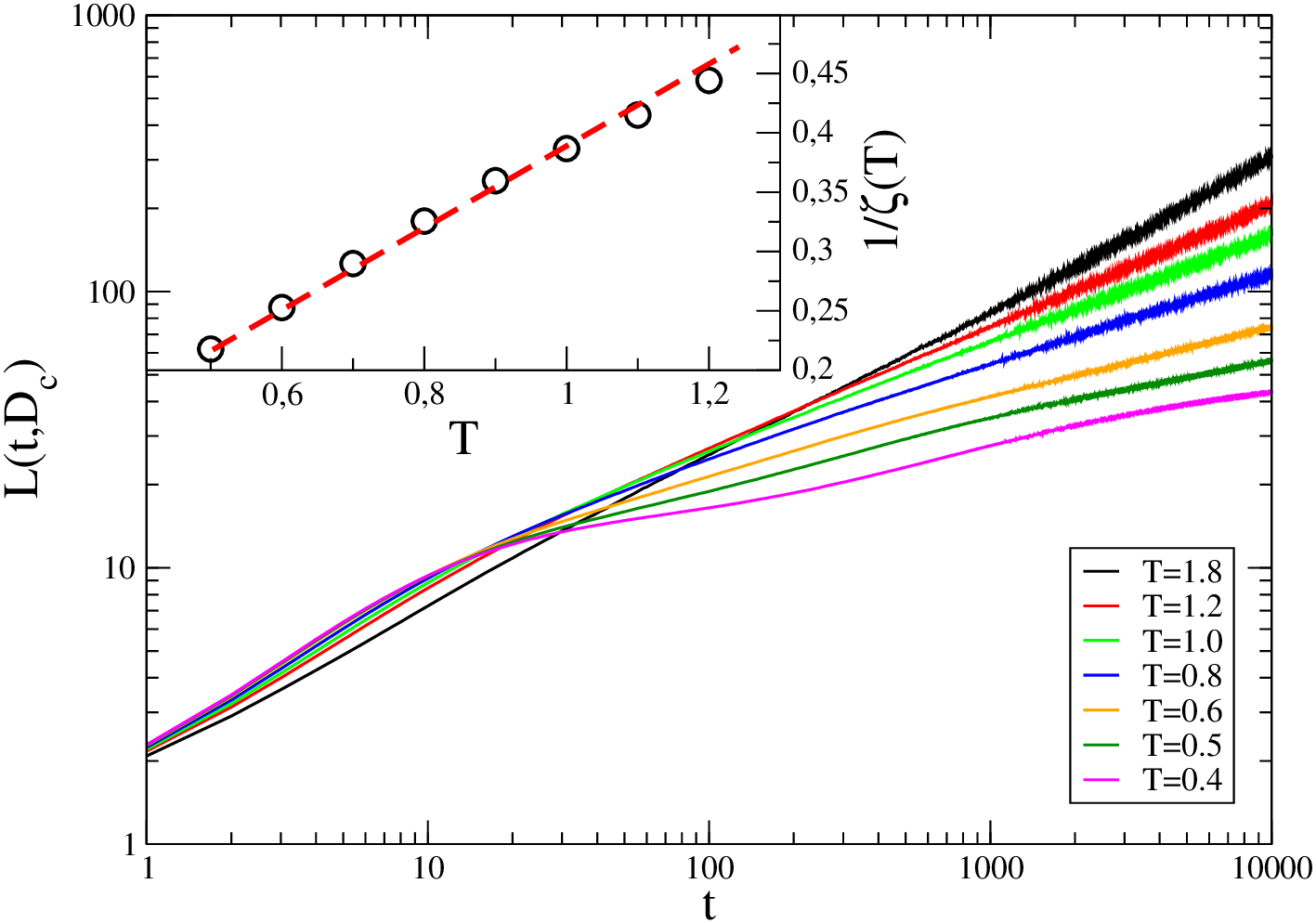}}}
      \caption{(Color online) The behavior of the typical length $L(t,D_c)$ after a quench to 
   different temperatures $T_{simul}$ (see key) of the SDIM is shown 
   in a double logarithmic plot. In the inset we plot the best fit exponent $1/\zeta(T)$ as function of $T$. The red dashed line is the theoretical value  (Eq.(\ref{expT})) with $A=0.36$.} 
\label{fig2}
\end{figure}

For $D=D_c$, on the other hand, the behavior of $L(t,D_c)$ is well fitted,
as expected, by a power-law $L(t,D_c)\simeq t^{1/\zeta (T)}$, with 
$1/\zeta (T_{simul})\simeq 0.41$. 
According to the discussion of Sec. \ref{over}, this behavior is 
expected irrespective of the quenching temperature (but with a different temperature-dependent
exponent $\zeta$), provided it
is sufficiently low. In order to check this we have performed a series 
of simulations by fixing $D=D_c$ and changing $T_{simul}$. The behavior of
$L(t,D)$ in these cases is shown in Fig. \ref{fig2}. The data are
consistent with power-laws for any choice of $T_{simul}$. Notice however that
for very small $T_{simul}$ the power-law is decorated by a sort of log-time
periodic oscillation. The exponent $1/\zeta (T)$, obtained by fitting all the curves
in the last decade (for $T=0.5$ we use two decades in order to smear out the oscillation), 
is plotted in the inset. According to Eq. (\ref{expT}),
one finds a linear behavior at sufficiently low temperatures. Fitting this 
curve we obtain $A\simeq 0.36$ (Eq.(\ref{expT})). Notice that for large values of the temperature this
linear behavior breaks down, due to the constraint $1/\zeta (T) \le 1/2$ 
since it is impossible to grow faster than in the pure case.

\subsubsection{3d results: BDIM}

\begin{figure}[t]
    \centering
    \vspace{1cm}                                       
   \rotatebox{0}{\resizebox{.65\textwidth}{!}{\includegraphics{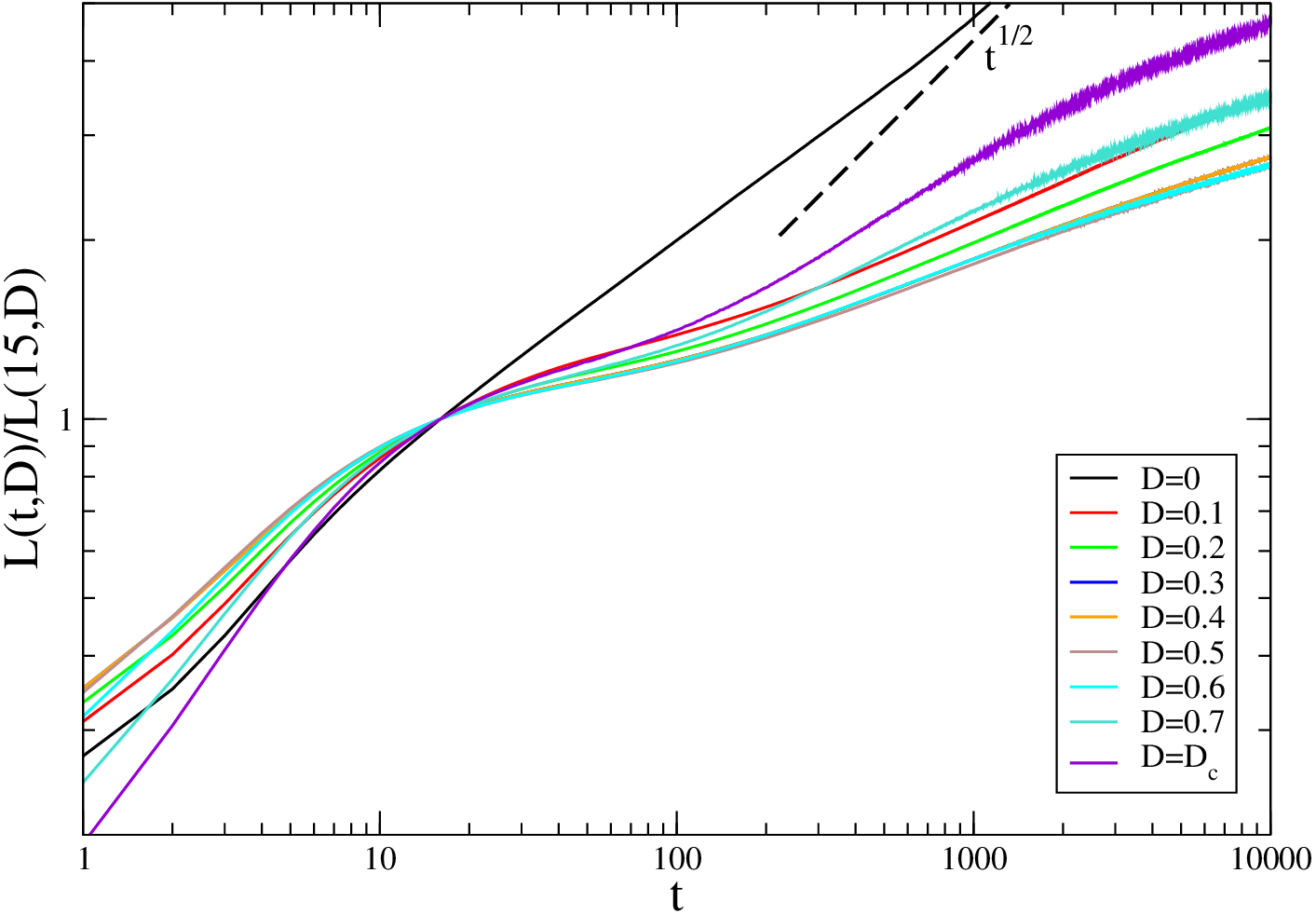}}}
      \caption{(Color online) The behavior of the typical length $L(t,D)/L(15,D)$ after a quench to 
$T=T_{simul}$ is shown for the BDIM 
   for several dilutions $D$ (see legend) in a double logarithmic plot. 
   The dashed-black line is pure growth-law $L(t,0)\sim t^{1/2}$.}
\label{fig3}
\end{figure}

In order to check the generality of the behaviors observed in the SDIM we have
performed a series of simulations also for the 3d-BDIM. Unfortunately, for this model
it is not possible to find a region of sufficiently low temperatures to be 
considered representative of the $T\to 0$ limit, as discussed in Sec. \ref{timev}, but
high enough to eliminate the stop and go behavior of the interfaces, which is responsible for 
the oscillatory character
of $L(t,D)$. We have explored several values of  $T_{simul}<1$, and in all cases we 
observe a non-monotonic behavior of $L(t,D)$ as $D$ is varied. In   Fig. (\ref{fig3}) we plot results for $T_{simul}=0.75$. For quenches at this temperature 
oscillations are clearly evident ruling out the possibility to make any statement on the form of the growth-law
(whether logarithmic or power-law). Despite this, the non monotonic behavior of $L(t,D)$ as
$D$ is varied at fixed $t$ is quite evident from Fig. \ref{fig3}.
Starting from the pure case $D=0$, where a fast growth is observed (although much slower than the
asymptotic expected behavior $L(t,0)\sim t^{1/2}$, due to the long-lasting pre-asymptotic
corrections discussed above) the growth gets slower as $D$ is increased up to $D=D^*\simeq 0.5$,
and then speeds up again until $D_c$ is reached. This shows that the 
non-monotonicity of the speed of growth as disorder is increased 
is a general property of diluted system, irrespective of dimension 
and of the kind of dilution. This suggests that an interpretation based on the
presence of three fixed points,
the pure one and those at $D^*$ and $D_c$, and the attractive/repulsive nature 
of such fixed points, can be appropriated quite generally.

\section{Conclusions} \label{CONCL}

In this paper we have studied the diluted Ising model in three dimensions.
The aim of the paper was to show the generality of the phase-ordering behavior
previously observed in the 2d-SDIM and 2d-BDIM, where it was first shown \cite{diluted,nuovo}
the non-monotonous dependence of the speed of growth on the amount of disorder, a fact
that was interpreted as due to
the close relation between the coarsening properties and some topological features 
of the network in disordered magnets. In order to do that we have considered three-dimensional systems.

Our results show that the pattern of behavior observed in these models is 
analogous to the one found in the 2d SDIM and RBIM, and henceforth an analogous
interpretation can be given. In the pure system phase-ordering occurs without
activation energy by a smooth curvature-driven displacement of interfaces,
leading to the well known growth-law $L(t,0)\sim t^{1/2}$. The situation changes
radically as soon as disorder is introduced, because inhomogeneities pin the
interfaces in energy minima and thermal activation is required to proceed.
In \cite{diluted} an argument was developed leading to the conjecture that
the $L(t,D)$-dependence of the pinning barriers changes when passing from 
$D<D_c$ to $D=D_c$, due to a radically different large-scale topology of the 
network which turns from a compact to a fractal object. This 
implies also a different growth-law, logarithmic for $D<D_c$ vs power-law
$L(t,D_c)\sim t^{1/\zeta(T)}$ right at $D_c$.  
In a renormalization-group language this features
can be accomodated in a framework with three fixed points, located
at $D=0$, $D=D^*$ and $D=D_c$.
Accordingly, a crossover phenomenon is 
observed where, for any $0<D<D_c$,
$L(t,D)$ changes from an early behavior associated to the nearest repulsive fixed 
point to the asymptotic logarithmic law. This gives rise to the non-monotonic
behavior of $L(t,D)$ as $D$ is varied keeping $t$ fixed. 

The presented numerical study suggests that the pattern of behaviors observed in \cite{diluted,nuovo} 
is quite a general feature of ferromagnetic disordered systems. Indeed we find that the non-monotonic behavior of $L(t,D)$ is observed 
also in $3d$ systems. Our results also indicate that, at least for the SDIM, $L(t,D_c)$ exhibits a power 
law growth  with a growth exponent $\zeta(t)$ inversely proportional to the temperature. 
The growth law becomes slower as $D$ is decreased to $D^*$, even if our data do not allows 
the precise determination of a logarithmic growth for $L(t,D^*)$. 
It would be an interesting line of research to investigate if a similar structure
underlies also the kinetics of other disordered systems such as, for instance,
Ising models with random fields.

{\bf Acknowledgments}
F.Corberi acknowledges financial support by MURST PRIN $2010HXAW77$\_$005$.


\begin{thebibliography}{99}

\bibitem{bray}
A.J. Bray, Adv. Phys. {\bf 43}, 357 (1994).

\bibitem{2dinoirf}
S. Puri, in {\it Kinetics of Phase Transitions}, edited by S. Puri and V. Wadhawan, CRC Press, Boca Raton (2009), p. 1.

\bibitem{zan} M.Zannetti, in {\it Kinetics of Phase Transitions} (Ref. \cite{2dinoirf}), p.153.

\bibitem{BCKM} J.P.Bouchaud, L.F.Cugliandolo, J.Kurchan and M.Mezard, in {\it Spin Glasses and Random Fields},
edited by A.P.Young (World Scientific, Singapore, 1997).

\bibitem{ioeleti}
F. Corberi, L. F. Cugliandolo, and H. Yoshino, 
in {\it Dynamical Heterogeneities in Glasses, Colloids, and Granular Media}, edited by L. Berthier, G. Biroli, 
J.-P. Bouchaud, L. Cipelletti, and W. Van Saarloos (Oxford University Press, Oxford, 2011).

\bibitem{rev}
F. Corberi, E. Lippiello and M. Zannetti, J. Stat. Mech.: Theory and Experiment P07002 (2007).

\bibitem{crp}
F. Corberi, Comptes rendus - Physique {\bf 16}, 332 (2015).

\bibitem{Puri}
S. Puri, Phase Transitions {\bf 77}, 469 (2004). 

\bibitem{diluted}
F. Corberi, E. Lippiello, A. Mukherjee, S. Puri, M. Zannetti,
Phys. Rev. E {\bf 88}, 042129 (2013).

\bibitem{nuovo}
F. Corberi, R. Burioni, E. Lippiello, A. Vezzani, M. Zannetti, 
Phys. Rev. E {\bf 91}, 062122 (2015).

\bibitem{puripowerlaw}
R. Paul, S. Puri and H. Rieger, Europhys. Lett. {\bf 68}, 881 (2004); 
R. Paul, S. Puri and H. Rieger, Phys. Rev. E {\bf 71}, 061109 (2005);
R. Paul, G. Schehr and H. Rieger, Phys. Rev. E {\bf 75}, 030104(R) (2007).

\bibitem{hp06}
M. Henkel and M. Pleimling, Europhys. Lett. {\bf 76}, 561 (2006); Phys. Rev. B {\bf 78}, 
224419 (2008).

\bibitem{lastparma}
R. Burioni, F. Corberi, and A. Vezzani, Phys. Rev. E {\bf 87}, 032160 (2013).

\bibitem{decandia} F. Corberi, A. de Candia, E. Lippiello and M. Zannetti, Phys. Rev. E
{\bf 65}, 046114 (2002);
F. Corberi, A. de Candia, E. Lippiello and M. Zannetti, Physica A {\bf  314}, 454 (2002).

\bibitem{EPL} E. Lippiello, A. Mukherjee, S. Puri and M. Zannetti, Europhys. Lett. {\bf 90}, 46006 (2010).

\bibitem{CLMPZ} F. Corberi, E. Lippiello, A. Mukherjee, S. Puri and M. Zannetti, 
J. Stat. Mech.: Theory and Experiment P03016 (2011).

\bibitem{variousnoi2}
F. Corberi, E. Lippiello, A. Mukherjee, S. Puri, and M. Zannetti,
Phys. Rev. E {\bf 85}, 021141 (2012).

\bibitem{various2}
S. Puri and N. Parekh, J. Phys. A {\bf 26}, 2777 (1993);
E. Oguz, A. Chakrabarti, R. Toral and J.D. Gunton, Phys. Rev. B {\bf 42}, 704 (1990);
E. Oguz, J. Phys. A {\bf 27}, 2985 (1994);
M. Rao and A. Chakrabarti, Phys. Rev. Lett. {\bf 71}, 3501 (1993);
C. Aron, C. Chamon, L.F. Cugliandolo and M. Picco, J. Stat. Mech. P05016 (2008). 
C. Castellano, F. Corberi, U. Marini Bettolo Marconi, and A. Petri,
J. Phys. IV France {\bf 08}, Pr6-93 (1998). 

\bibitem{10noirf}
L.F. Cugliandolo, Physica A {\bf 389}, 4360 (2010).

\bibitem{pp} H. Park and M. Pleimling, Phys. Rev. B {\bf 82}, 144406 (2010).

\bibitem{HH} D.A. Huse and C.L. Henley, Phys. Rev. Lett. {\bf 54}, 2708 (1985).

\bibitem{parma}
R. Burioni, F. Corberi, and A. Vezzani, 
J. Stat. Mech. (2010) P12024; J. Stat. Mech. (2009) P02040;
R. Burioni, D. Cassi, F. Corberi, and A. Vezzani, Phys. Rev. E {\bf 75}, 011113 (2007);
Phys. Rev. Lett. {\bf 96}, 235701 (2006).

\bibitem{pcp91}
S. Puri, D. Chowdhury and N. Parekh, J. Phys. A {\bf 24}, L1087 (1991);
S. Puri and N. Parekh, J. Phys. A {\bf 25}, 4127 (1992).

\bibitem{bh91}
A.J. Bray and K. Humayun, J. Phys. A {\bf 24}, L1185 (1991).

\bibitem{sab08}
A. Sicilia, J. J. Arenzon, A. J. Bray and L. F. Cugliandolo, Europhys. Lett. {\bf 82}, 1001 (2008);
M. P. O. Loureiro, J. J. Arenzon, L. F. Cugliandolo, and A. Sicilia, Phys. Rev. E {\bf 81}, 021129 (2010).

\bibitem{nota1} As conjectured in \cite{lastparma}, 
a power-law growth of $L(t)$ with a
temperature dependent exponent is expected also, besides on a percolation network, 
on a certain class of fractal structures.

\bibitem{Stauffer}
D. Stauffer, Phys. Repts. {\bf 54}, 1 (1979);
D. Stauffer and A. Aharony, {\it Introduction to Percolation Theory}, Taylor and Francis, London 1994 (revised second edition).

\bibitem{lip}
E. Lippiello, and M.Zannetti, Phys. Rev. E {\bf 61}, 03369 (2000);

\bibitem{Claudio}
F. Corberi, C. Castellano, E. Lippiello, and M. Zannetti, Phys. Rev. E {\bf 65}, 066114 (2002);
N. Andrenacci, F. Corberi, and E. Lippiello, Phys. Rev. E {\bf 74},
031111 (2006);
E. Lippiello, F. Corberi, and M. Zannetti, Phys. Rev. E {\bf 71}, 036104 (2005);
R. Burioni, F.Corberi, and A. Vezzani, Phys. Rev. E {\bf 79}, 041119 (2009);
S. J. Cornell, K. Kaski, and R. B. Stinchcombe, Phys. Rev. B {\bf 44}, 12263 (1991).

\bibitem{nobulk}
F. Corberi, E. Lippiello, and M. Zannetti, Phys. Rev. E {\bf 63}, 061506 (2001);
Eur. Phys. J. B {\bf 24} (2001), 359; Phys.Rev. E {\bf 68}, 046131 (2003); Phys.Rev. E {\bf 78}, 011109 (2008);
E. Lippiello, F. Corberi, A. Sarracino, and M. Zannetti, Phys. Rev. E {\bf 78}, 041120 (2008);
F. Corberi, and L.F. Cugliandolo, J. Stat. Mech. P05010 (2009). 

\bibitem{altriIsing}
J.G. Amar and F. Family, Bull. Am. Phys. Soc. {\bf 34}, 491 (1989); J.D. Shore, M. Holzer and J.P. Sethna, Phys. Rev. B {\bf 46},
11376 (1992).

\bibitem{noiIsing}

F. Corberi, E. Lippiello, M. Zannetti, Phys. Rev. E {\bf 78}, 011109 (2008);



\end{thebibliography}
\end{document}